\documentclass[journal]{IEEEtran}
\usepackage[utf8]{inputenc}
\usepackage{algorithm2e}

\usepackage{cite}

 \usepackage{url}
\usepackage{footnote}

\usepackage{dsfont}
\ifCLASSINFOpdf
  \usepackage[pdftex]{graphicx}

\else

\fi

\usepackage[cmex10]{amsmath}
\usepackage{amsfonts}
\usepackage{color}
\usepackage[space]{grffile}
\usepackage[]{diagbox}
\usepackage{amsmath,graphicx}
\usepackage[utf8]{inputenc}
\usepackage{amsmath,amssymb}
\usepackage{mathrsfs}
\usepackage{dsfont}
\usepackage{bm}

\begin{document}
%
\title{Spectral Unmixing: A Derivation of the Extended Linear Mixing Model from the Hapke Model}
%
%
%
\author{Lucas~Drumetz,~\IEEEmembership{Member,~IEEE}  
        Jocelyn~Chanussot,~\IEEEmembership{Fellow,~IEEE}        
         and~Christian~Jutten,~\IEEEmembership{Fellow,~IEEE,}
\thanks{L. Drumetz is with IMT Atlantique, Lab-STICC, UBL, Technopôle Brest-Iroise CS 83818, 29238 Brest Cedex 3, France (e-mail: lucas.drumetz@imt-atlantique.fr)}\thanks{J. Chanussot and C. Jutten are Univ. Grenoble Alpes, CNRS, Grenoble INP, GIPSA-lab, 38000 Grenoble, France.  (e-mail: \{jocelyn.chanussot,christian.jutten\}@gipsa-lab.grenoble-inp.fr).}
\thanks{This work has been supported by the 2012 ERC Advanced Grant project CHESS (Grant \# 320684), as well as the  project ANR-DGA APHYPIS, under grant ANR-16 ASTR-0027-01.}}
\maketitle
\begin{abstract}
In hyperspectral imaging, spectral unmixing aims at decomposing the image into a set of reference spectral signatures corresponding to the materials present in the observed scene and their relative proportions in every pixel. While a linear mixing model was used for a long time, the complex nature of the physical mixing processes led to shift the community's attention towards nonlinear models or algorithms accounting for the variability of the endmembers. Such intra-class variations are due to local changes in the physico-chemical composition of the materials, and to illumination changes. In the physical remote sensing community, a popular model accounting for illumination variability is the radiative transfer model proposed by Hapke. It is however too complex to be directly used in hyperspectral unmixing in a tractable way. Instead, the Extended Linear Mixing Model (ELMM) allows to easily unmix hyperspectral data accounting for changing illumination conditions and to address nonlinear effects to some extent. In this letter, we show that the ELMM can be obtained from the Hapke model by successive simplifiying physical assumptions, whose validity we experimentally examine, thus demonstrating its relevance to handle illumination induced variability in the unmixing problem.
\end{abstract}
\begin{IEEEkeywords}
Hyperspectral image unmixing, spectral variability, Hapke model, extended linear mixing model.
\end{IEEEkeywords}
%
\IEEEpeerreviewmaketitle
\section{Introduction}
\IEEEPARstart{H}{yperspectral} imaging provides information in (typically) hundreds of wavelengths in the visible and near infrared domains of the electromagnetic spectrum. The spectral resolution is then much finer than that of color or multispectral images. However, the spatial resolution is conversely coarser. Several distinct materials can then be present in the field of view of a single pixel. The observations captured at the sensor level are then mixtures of the contribution of each material. The inverse problem, called unmixing, aims at identifying the spectra of the pure materials present in the scene (called \emph{endmembers}) and to estimate their relative proportions in each pixel (called \emph{fractional abundances}).

Usually, a linear mixing model (LMM) models the relationship between the observed data, the endmembers and their abundances~\cite{Bioucas2012} and writes $\mathbf{X} = \mathbf{SA} + \mathbf{E}$. The image is represented as a matrix $\mathbf{X}\in\mathbb{R}^{L\times N}$, where $L$ is the number of spectral bands, and $N$ is the number of pixels. The endmembers $\mathbf{s}_{p}$, $p = 1,...,P$ are gathered in the columns of a matrix $\mathbf{S}\in \mathbb{R}^{L\times P}$, where $P$ is the number of materials. The abundance coefficients for each pixel and each material are stored in a matrix $\mathbf{A}\in \mathbb{R}^{P\times N}$, and $\mathbf{E}$ is an additive noise. The abundances are proportions, usually constrained to be positive, and to sum to one in each pixel. Geometrically, the LMM constrains the data to lie in a simplex spanned by the endmembers. In many cases, the LMM is a reasonable approximation of the physics of the mixtures.

Nevertheless, a key limitation is related to possible nonlinear mixing phenomena, e.g. in urban scenarios or tree canopies, when light bounces on several materials before reaching the sensor. Another situation is intimate mixing phenomena in particulate media~\cite{Heylen2014}.

\begin{figure}
\begin{center}
\includegraphics[scale=0.37]{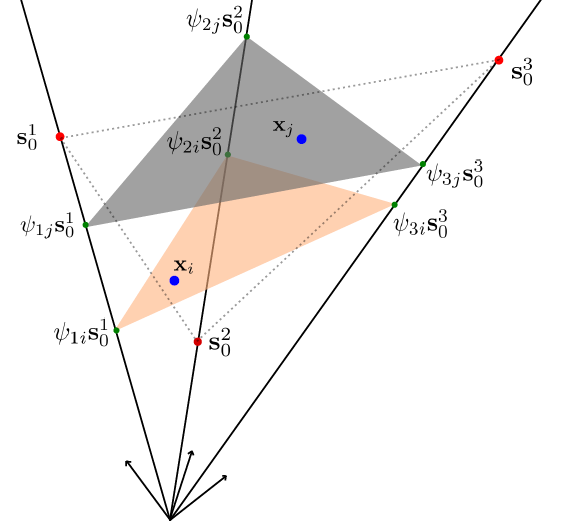}
\caption{Geometric interpretation of the ELMM in the case of three endmembers~\cite{drumetztip}. In blue are two data points, in red are the reference endmembers and in green are the scaled versions for the two considered pixels. The simplex used in the LMM is shown in dashed lines.}
\label{model}
\end{center}
\end{figure}

The other important limitation, if not predominant, comes from the underlying assumption in the LMM that each endmember is explained by a unique spectral signature. This is a convenient approximation, but an endmember is actually more accurately described by a collection of signatures, which account for the intra-class variability of that material~\cite{zare2014,drumetz2016endmember}. Spectral variability approaches of the literature essentially boil down to being able to estimate variants of the materials' spectra in each pixel. Many physical phenomena can induce variations on the spectra of pure materials, be it a change in their physico-chemical composition, or the topography of the scene, which locally changes the incidence angle of the light and the viewing angle of the sensor. This phenomenon is referred to as \emph{endmember variability}~\cite{PLMM,halimi2015,henrot2016}. A physics-inspired model to explain illumination induced variability is the Extended Linear Mixing Model (ELMM)~\cite{drumetztip}:
\begin{equation}
\mathbf{x}_{n} = \sum_{p=1}^{P} a_{pn}\psi_{pn} \mathbf{s}_{0p} + \mathbf{e}_{n}
\label{var}
\end{equation}
where $a_{pn}$ is the abundance coefficient for material $p$ and pixel $n$, and $\mathbf{e}_{n}$ is an additive noise. $\psi_{pn}$ is a positive scaling factor whose effect is to rescale locally each endmember, and $\mathbf{s}_{0p}$ is a reference endmember for material $p$. This model can be empirically validated by many experimental measurements of spectra of the same materials, whose shapes remains the same but whose amplitudes vary according to a scaling (see e.g. Fig. 1 of~\cite{drumetztip}). The ELMM enjoys a simple geometric interpretation: the data lie in a convex cone spanned by the reference endmembers, which define lines on which the local endmembers can lie. Each pixel belongs to a simplex spanned by the local endmembers (Fig.~\ref{model}).
This model has been used in several works since it was introduced, e.g. in ~\cite{halimi2017fast,hong2018sulora,borsoi2018super,uezato2019hyperspectral}. Variations in the spectra of one given material
due to changing illumination conditions experimentally
appear to be reasonably well explained by a scaling variation. Another complex physical semi-empirical model was introduced by Hapke to model both intimate mixing phenomena and reflectance variations induced by the changing geometry of the scene~\cite{Hapke2012}.

\begin{figure}
\begin{center}
\vspace{-0.5cm}
\raisebox{-0.5\height}{\includegraphics[scale=0.22]{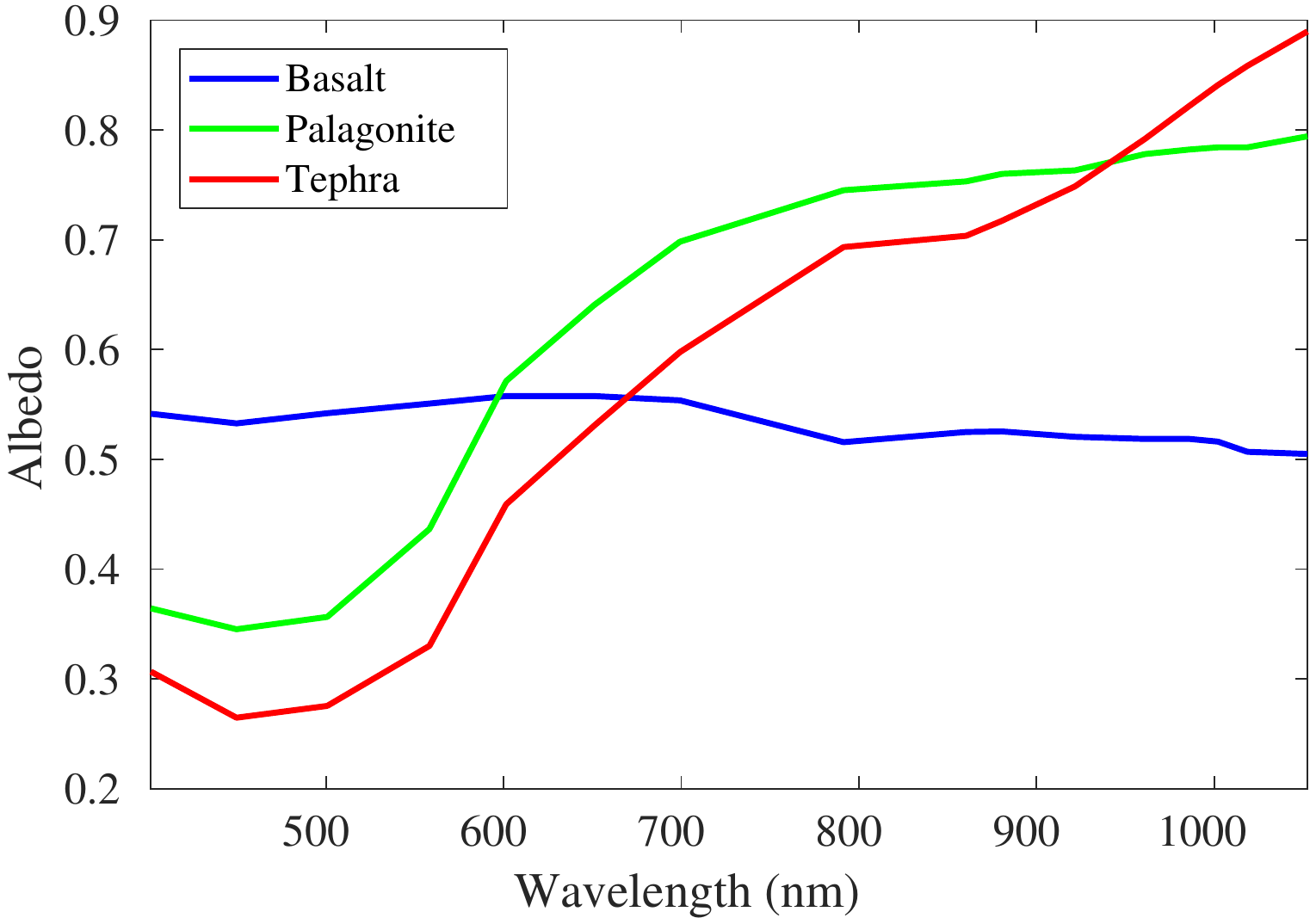}}
\raisebox{-0.5\height}{\includegraphics[scale=0.11]{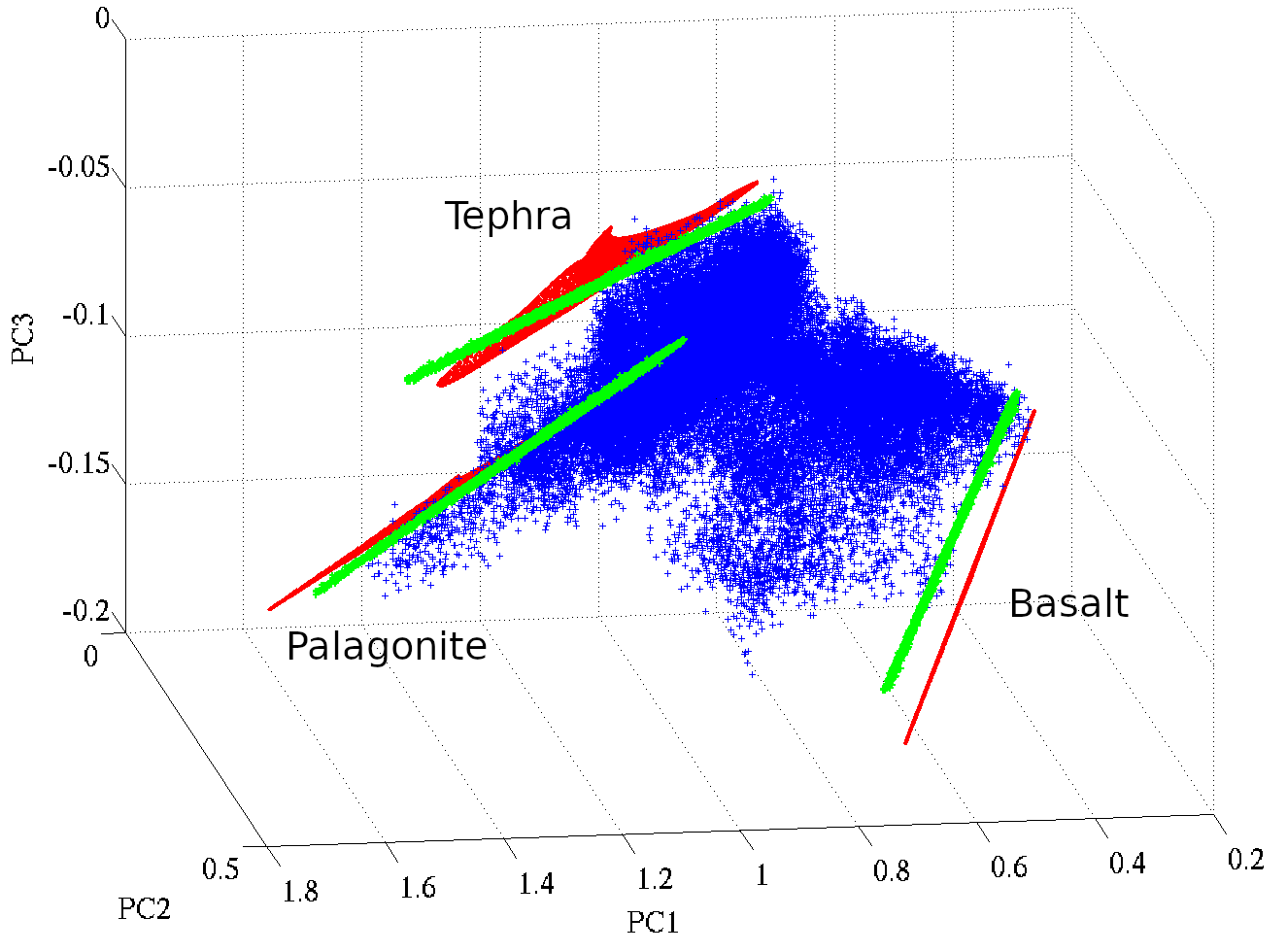}}
\caption{Left: Albedo endmembers of three materials. Right: scatterplot of the first three components of a PCA of a dataset simulated with the LMM (blue), using the endmember variants generated by the Hapke model (red). The endmembers estimated by the ELMM are in green~\cite{drumetztip}.}
\label{pca2}
\end{center}
\end{figure}

The ELMM was experimentally shown to better fit Hapke simulated data than other approaches tackling endmember variability in unmixing. These results are presented in detail in~\cite{drumetztip}. We reproduce in Fig.~\ref{pca2} a figure showing a linearly mixed dataset with Hapke generated endmembers (red), their approximations by the ELMM (green). The materials of interest are commonly found on small bodies of the solar system. Basalt has a relatively low and flat spectrum, and thus is less affected by the nonlinearities of the Hapke model than tephra and palagonite. In any case, the ELMM provides a very good approximation of the red manifolds generated by the Hapke model. For experiments on the capability of the ELMM to explain variability with real data acquired in various contexts, we refer e.g. to~\cite{drumetztip,drumetzICASSP2018,ibarrola2019hyperspectral}.

Interestingly, the ELMM is not strictly speaking a linear mixing model, because of the pixel and endmember dependent scaling factors. Besides, those scaling factors were actually proven to be able to capture nonlinear mixing effects in a satisfactory way in a previous study~\cite{drumetz2017relationships}.

In this letter, for the first time, we prove how the ELMM can be derived from the Hapke model by several successive approximations. In addition, the derivation and experiments give insight on the ELMM, by showing when it approximates the Hapke model accurately for small albedos, and certain favorable geometrical configurations, confirming past experimental results~\cite{drumetztip}.

The remainder of this letter is organized as follows: section~\ref{Hapke} briefly introduces the Hapke model and its parameters, section~\ref{link} derives the ELMM from the Hapke model, section \ref{exp} provides further experimental insight on the approximations, and finally, section~\ref{concl} gathers a few concluding remarks.
\section{The Hapke Model}
\label{Hapke}
Here, we briefly describe the Hapke model and how it models reflectance as a function of various physical parameters.
The complete analytical expressions of all the terms involved, and let alone a detailed derivation of the Hapke model are far beyond the scope of this letter. We refer to~\cite{Hapke2012} for the original derivation, and e.g. to~\cite{Heylen2014} for a gentle introduction to the model. From an unmixing point of view, this model is too complex as is (for example, it is non injective) and its physical parameters are not available in practice. 

Reflectance, the physical quantity usually used to work with hyperspectral remote sensing images (after atmospheric correction of radiance units), is dependent on the geometry of the acquisition. Depending on the incidence and viewing angles, the measured reflectance can significantly differ. The reflectance of a material is also influenced by its photometry, i.e. the way light interacts with the material. Photometry can be modeled through some optical parameters (surface roughness, scattering behavior...) of the materials. We will briefly describe the photometric parameters involved in the model, but for a more thorough description of their physical and geological interpretations we refer to~\cite{fernando2015characterization}. The albedo of material, contrary to its reflectance, is truly characteristical of the material and depends neither on the geometry of the scene nor on the photometry of the considered material.

The Hapke model is essentially an equation providing the bidirectional reflectance in a given wavelength of a material as a function of its albedo for that wavelength, and of parameters defining the geometry of the acquisition and characterizing the photometry of the observed material. We assume the mixture of the materials occurs at the macroscopic level, and hence we do not consider intimate mixing, which can also be explained by Hapke's model. Therefore, the LMM assumption remains approximately valid in each pixel. The equations below are to be understood to be applied separately to each endmember, each pixel and each wavelength of a hyperspectral image, using its pure albedo spectrum, the photometry and the local geometry. This defines local reflectance endmember variants in each pixel, which are then linearly mixed.

\begin{figure}
\centering
\includegraphics[scale=0.19]{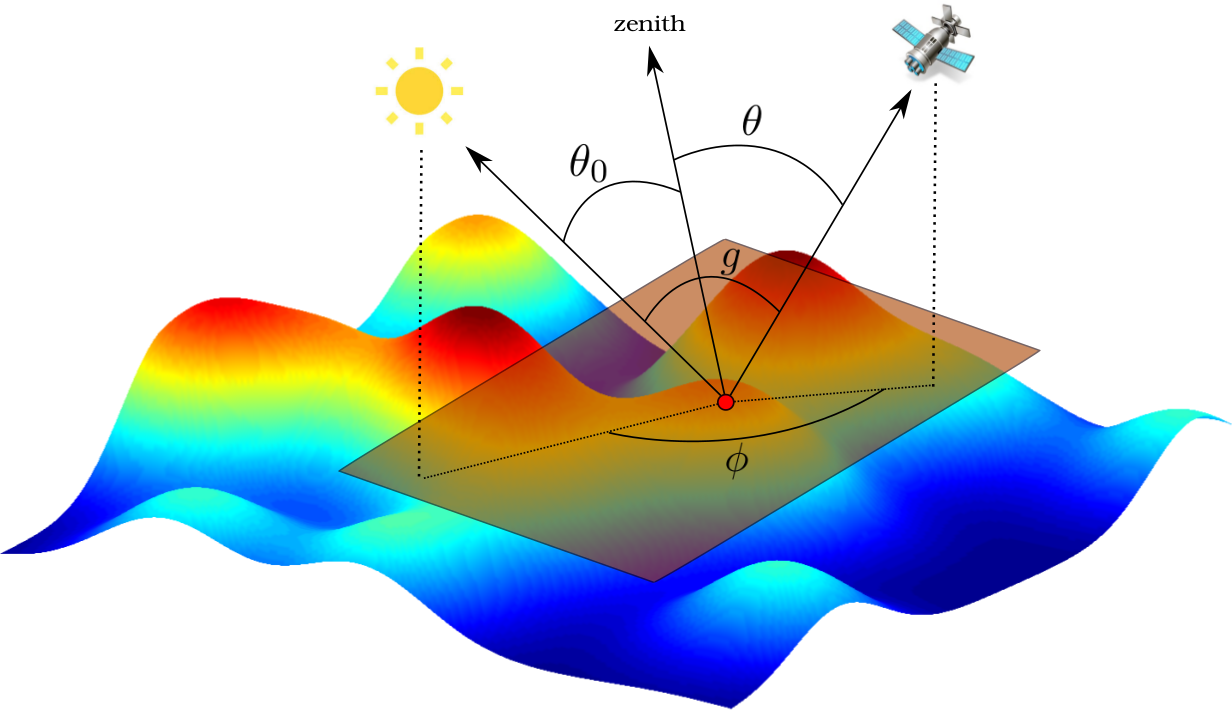}
\caption{Acquisition angles for a given spatial location (red dot). The tangent plane at this point of the surface is in brown. The incidence angle is $\theta_{0}$, the emergence angle is $\theta$, and the angle between the projections of the sun and the sensor is the azimuthal angle, denoted as $\phi$. $g$ is the phase angle. $\theta_{0}$ and $\theta$ are defined with respect to the normal to the surface at this point.}
\label{angles_soa}
\end{figure}

The local pixelwise geometry of the scene can be described by several parameters (Fig.~\ref{angles_soa})~\cite{Hapke2012,cord2005experimental}.
The zenith is locally defined as the direction of the normal vector to the tangent plane to the surface observed. Depending on the topography, this plane is different for each pixel. The angle between the zenith and the sun is called the sun zenith angle, or incidence angle, $\theta_{0}$. The angle between the zenith and the sensor is called the emergence angle, $\theta$. The angle between the sun and sensor directions (with the origin on the field of view of the current pixel) is called the phase angle $g$. Finally, the angle between the projections of the sun and the sensor on the tangent plane is called the azimuthal angle $\phi$. These four angles completely characterize the geometry of a pixel's acquisition. Hapke's model can be expressed as~\cite{Hapke2012,Heylen2014}:
\begin{multline}
\rho(\omega,\mu,\mu_{0},\phi,g) = \frac{\omega}{4(\mu_{e} + \mu_{0e})} S(\mu,\mu_{0},\phi)   \times   \\
 ((1 + B(g))P(g) + H(\omega,\mu_{e})H(\omega,\mu_{0e}) - 1),
\label{full_hapke}
\end{multline}
$\rho$ is the reflectance for a given wavelength range, $\mu = \textrm{cos}(\theta)$, $\mu_{0} = \textrm{cos}(\theta_{0})$, $\omega$ is the single scattering albedo of the material for the same wavelength range, $P$ is the phase function, modeling the angular scattering distribution of the material:
\begin{equation}
P(g) = \frac{c(1-b)^2}{(1-2b\cos(g) + b^2)^{3/2}} + \frac{(1-c)(1-b)^2}{(1+2b\cos(g) + b^2)^{3/2}}.
\end{equation}
$B$ is a function related to the opposition effect (brightening of the observed surface when the illumination comes from behind the sensor, i.e. for small $g$ values):
\begin{equation}
B(g) = \frac{B_{0}}{1+(1/h) \tan (g/2)},
\end{equation}
and $H$ is the isotropic multiple scattering function: 
\begin{equation}
H(\omega,\mu) \approx \frac{1+ 2\mu}{1+2\mu(\sqrt{1-\omega})},
\label{scattering}
\end{equation}
$\mu_{0e}$ and $\mu_{e}$ are the cosines of the modified incidence and emergence angles, accounting for the macroscopic roughness of the materials. $S(\mu,\mu_{0},\phi)$ is a \emph{shadowing} function, reducing the total reflectance when surface roughness hides parts of the observed surface from the sensor, or shadows a fraction of it. In the remainder of the paper, we will assume that the surface of the materials is smooth, so that there is no shadowing effect, and the emergence and incidence angles are not modified, leading to $S(\mu_{0}, \mu,\phi) = 1$ and $\mu_{0e} = \mu_{0}$, and $\mu_{e} = \mu$.

$B$ and $P$ are parametrized by photometric parameters of the considered material. For the phase function $P$, the photometric parameters used are i) the asymmetry parameter of the scattering lobes $b$ ($0 \leq b \leq 1 $), higher values meaning narrower lobes and higher scattering intensity, ii) the backward scattering fraction $c$ ($0 \leq c \leq 1 $); $c < 0.5$ means that the material mainly backscatters the incoming light towards the incidence direction, and $c > 0.5$ means that the material has a predominantly forward scattering behavior. As examples of particular behaviors of the phase function, we can cite specular reflection, characterized by $b = 1$ and $c = 1$, or Lambertian (isotropic) scattering, characterized by $b = 0$ and $c = 0.5$. For $B$, the parameters $h$ and $B_{0}$, account for the angular width and the strength of the opposition effect, respectively.
\section{Derivation of the ELMM}
\label{link}
\subsection{Simplifying the Hapke model}
Here, using simplifying assumptions, we go from the general Hapke model~(\ref{full_hapke}) to a special case of the ELMM presented in~\cite{Veganzones2014_ELMM,drumetztip}.

As explained in~\cite{Heylen2014}, assuming a Lambertian scattering, the phase function reduces to $P(g) = 1$. Besides, for Lambertian surfaces, there is no opposition surge ($B_{0} = 0$) since the scattering is isotropic. In any case, even for non Lambertian photometries, for large enough $g$, the opposition effect is negligible and $B(g) \approx 0$ anyway. 

Incorporating all these assumptions, and plugging the expression of the scattering function~(\ref{scattering}) in~(\ref{full_hapke}), the (bidirectional) reflectance becomes~\cite{Heylen2014}:
\begin{equation}
\rho(\omega,\mu,\mu_{0}) = \frac{(1+2 \mu)(1+2 \mu_{0}) \omega}{4(\mu + \mu_0)(1+2\mu \sqrt{1-\omega})(1+2\mu_{0} \sqrt{1-\omega})}.
\label{simple_hapke_init}
\end{equation}
Finally, we obtain the relative bidirectional reflectance by dividing by a reference value where $\omega = 1$, in which case $\rho(\omega = 1,\mu,\mu_{0}) = \frac{(1+2 \mu)(1+2 \mu_{0})}{4(\mu + \mu_0)} $. The reflectance $\rho_{0}$ is then:
\begin{equation}
\rho_0(\omega,\mu,\mu_{0}) = \frac{\omega}{(1+2\mu \sqrt{1-\omega})(1+2\mu_{0} \sqrt{1-\omega})}.
\label{simple_hapke}
\end{equation}
All the photometric effects are eliminated because of the Lambertian photometry assumption. The model is still material dependent, because the albedo spectrum depends on the material. The only other parameters left are geometry related parameters. However, the albedo spectrum is not available in practice. A workaround for this is to numerically invert the model (the full model for a more precise estimate) if all the parameters but the albedo are known in a pixel. In such a case, the reflectance-albedo relation is bijective. However, there is no simple way to assess the results of this method in practice, especially in real scenarios, and the incertitudes on the results could be very important. The principles of this strategy are applied to controlled lab measurements in~\cite{marrerovalidation}. Even then, the model is still complex, highly nonlinear, especially for high albedos, and it is not identifiable when no parameters are known, due to the symmetry of~(\ref{simple_hapke}) w.r.t. $\mu$ and $\mu_{0}$. 

For small single scattering albedo values (say up to 0.5),~(\ref{simple_hapke}) is close to linear, while important nonlinearities appear for large albedo values. The validity of the linear approximation actually also depends on the values of the incidence and emergence angles. In Fig.~\ref{albedos}, the function defined by~(\ref{simple_hapke}) is plotted (blue curves) for three values of the acquisition angles: when both the sensor and the sun are at nadir (Fig.~\ref{albedos} (a)), when the sensor and the sun both make an angle of 45 degrees with respect to the normal to the surface (Fig.~\ref{albedos} (b)), and with raking incident light ($\theta_{0} = 90^{\circ}$), and the sensor making an angle of 45 degrees with the nadir direction (Fig.~\ref{albedos} (c)). If both angles are equal to 90 degrees (with respect to the normal), the resulting reflectance equals the albedo. Here, because of these considerations, we propose to further approximate the relationship between albedo and reflectance by performing a first order Taylor expansion around $\omega =0$, with the angles fixed (in practice approximately valid for ``small" albedos):
\begin{figure}
\begin{center}
\begin{minipage}{0.32\linewidth}
  \centering
\includegraphics[width = 1\linewidth]{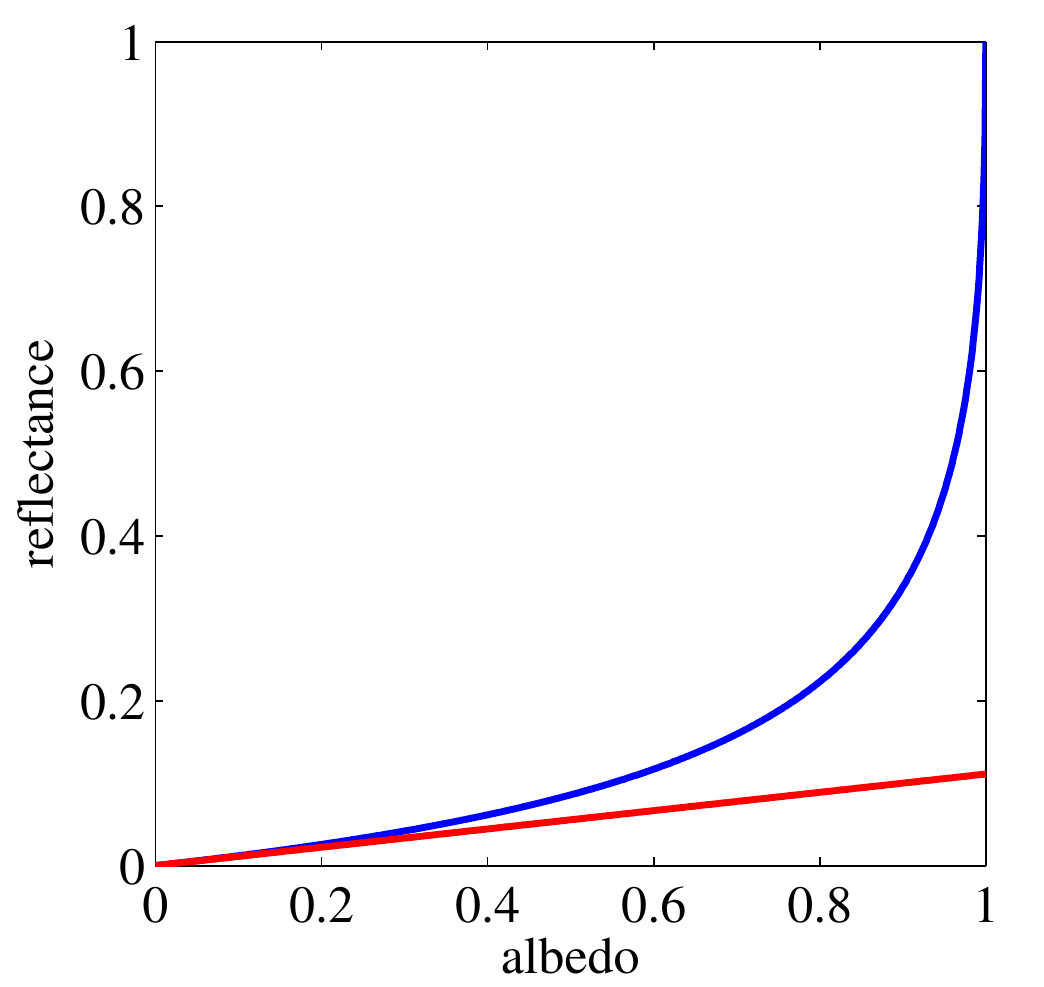}
  \centerline{\scriptsize{(a) $\theta_{0} = \theta = 0^{\circ}$}}\medskip
  \end{minipage}
\begin{minipage}{0.32\linewidth}
  \centering
\includegraphics[width = 1\linewidth]{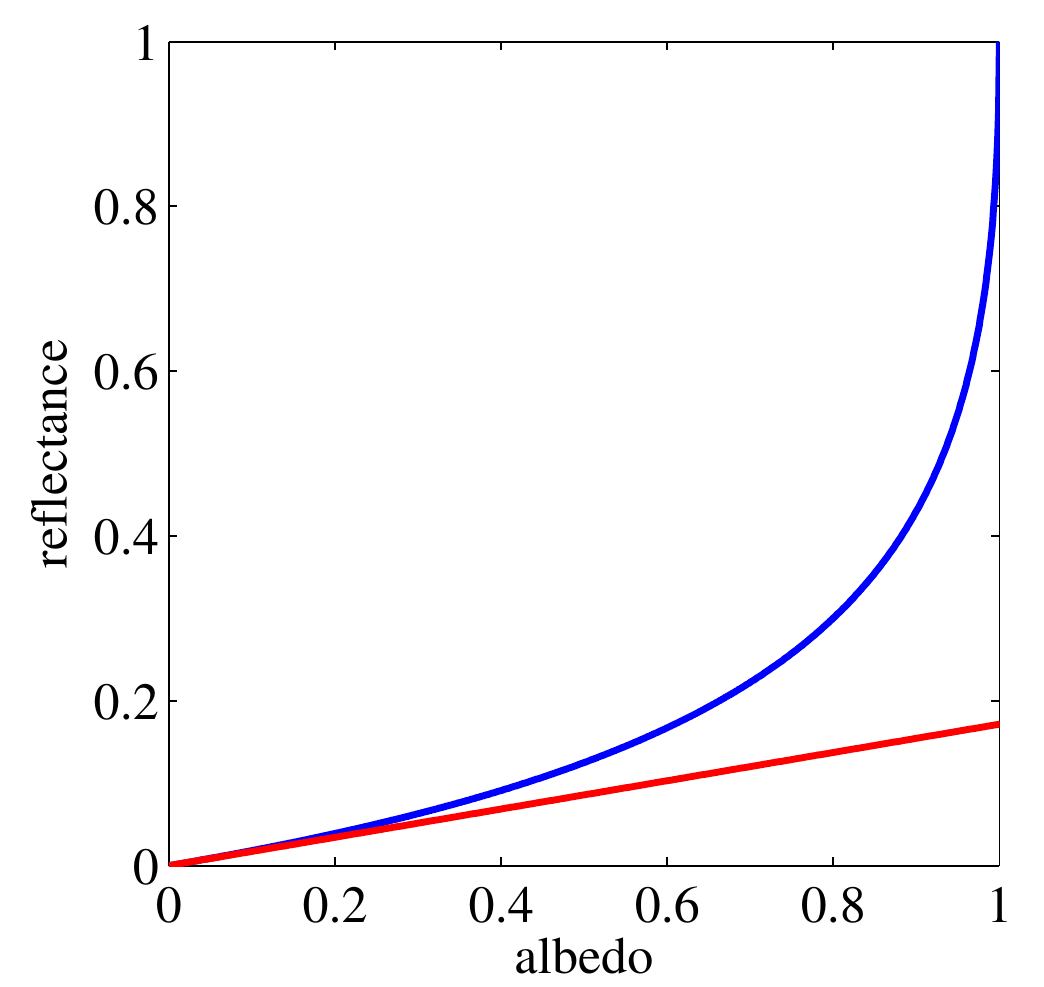}
  \centerline{\scriptsize{(b) $\theta_{0} = \theta = 45^{\circ}$}}\medskip
\end{minipage}
\begin{minipage}{0.32\linewidth}
  \centering
  \hspace{0.5cm}
\includegraphics[width = 0.975\linewidth]{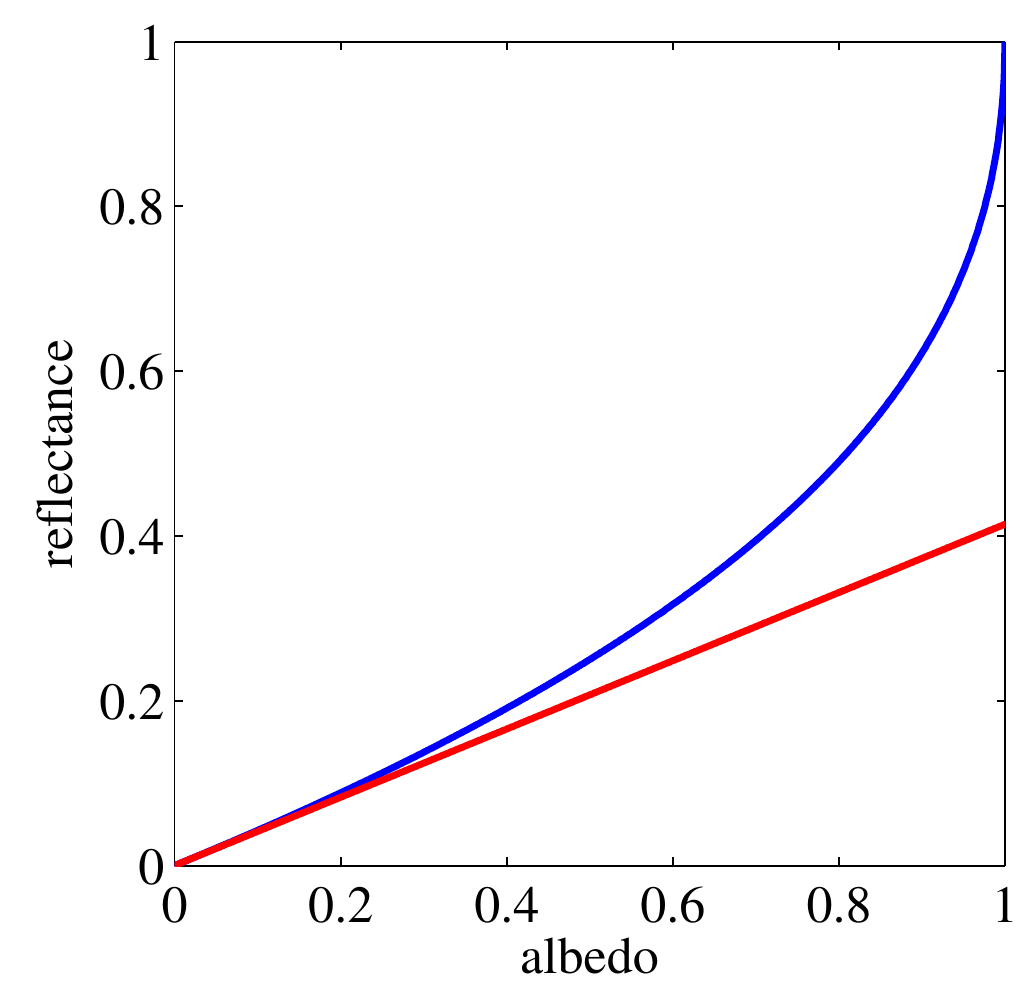}
  \centerline{\scriptsize{(c) $\theta_{0} = 90^{\circ}$, $\theta = 45^{\circ}$}}\medskip
\end{minipage}
\caption{Reflectance plotted as a function of the albedo according to~(\ref{simple_hapke}) (blue), and the  Taylor expansion~(\ref{Taylor}) in $\omega = 0$ (red), for three geometries.}
\label{albedos}
\vspace{-0.75cm}
\end{center}
\end{figure}
\begin{align}
\rho_0(\omega,\mu,\mu_{0}) &  = \rho_0(0) + \frac{\partial \rho_0}{\partial \omega}(0) \omega  + o (\omega) \nonumber \\
& = \frac{\omega}{4\mu \mu_{0} + 2 \mu + 2 \mu_{0} + 1}  + o (\omega).
\label{Taylor}
\end{align}
The coefficient of the expansion only depends on the geometry of the acquisition: it affects an albedo spectrum in the same way for any wavelength. Now let us assume that for a given material $p$, we have at our disposal a reference endmember $\mathbf{s}_{0p}$ (usually extracted from the data), with a geometry defined by the angles $\theta$ and $\theta_{0}$. This endmember is a collection of reflectances for various wavelengths. Then with the first order model~(\ref{Taylor}), the ratio between the reflectances in each wavelength is constant, and for the representative $\mathbf{s}_{pn}$ of endmember $p$ in pixel $n$ and small albedos:
\begin{equation}
\mathbf{s}_{pn} \approx \frac{4\mu_{n} \mu_{n0} + 2 \mu_{n} + 2 \mu_{n0} + 1}{4\mu \mu_{0} + 2 \mu + 2 \mu_{0} + 1} \ \mathbf{s}_{0p} = \psi_{n} \mathbf{s}_{0p}.
\label{scaling}
\end{equation}
From this equation, we see that now the link between the local representative of an endmember in a pixel and a reference signature for this material is a positive scaling factor incorporating the information about the geometry in the considered pixel. With this approximation, we make the connection between the semi-empirical model of Hapke and the well known fact in the remote sensing community that illumination effects can be well approximated by scaling variations of the spectra.

\subsection{ELMM description}
The considerations of the previous sections lead to plug this variability model to the usual LMM, so that it becomes:
\begin{equation}
\mathbf{x}_{n} = \psi_{n} \sum_{p=1}^{P} a_{pn} \mathbf{s}_{0p} + \mathbf{e}_{n} = \mathbf{S}_{0}  \psi_{n} \mathbf{a}_{n}  + \mathbf{e}_{n}.
\label{onepsi}
\end{equation} 
The LMM is simply scaled in each pixel by a different nonnegative scaling factor. In practice, it can be useful to allow the scaling factor to vary for each material:
\begin{equation}
\mathbf{x}_{n} = \sum_{p = 1}^{P} a_{pn} \psi_{pn}\mathbf{s}_{0p} + \mathbf{e}_{n} = \mathbf{S}_{0} \boldsymbol{\psi}_{n} \mathbf{a}_{n} + \mathbf{e}_{n},
\label{ELMM_scaling}
\end{equation}
where the $\psi_{pn}$ are now pixel and material dependent scaling factors, $\boldsymbol{\psi}_{n} \in \mathbb{R}^{P\times P}$ is a diagonal matrix, containing the scaling factors for each material on its diagonal. We thus recover the model of~(\ref{var}). The scaling factors can also be rearranged into a matrix $\boldsymbol{\Psi}\in \mathbb{R}^{P \times N}$ (the $n^\textsuperscript{th}$ column contains the diagonal of $\boldsymbol{\psi}_{n}$). This allows model~(\ref{ELMM_scaling}) to be rewritten globally for the whole image as $\mathbf{X} = \mathbf{S}_{0}(\mathbf{\Psi} \odot \mathbf{A}) + \mathbf{E}$,
with $\odot$ the Hadamard product.
The main reason behind the introduction of a scaling factor for each pixel and material is that it will make the model more flexible, allowing to model material dependent variabilities, e.g. related to material dependent photometric phenomena or more pragmatically to the intrinsic variability of each material. Another important reason is that this version of the ELMM was also proven theorically and experimentally in~\cite{drumetz2017relationships} to locally approximate nonlinear mixtures by absorbing potential nonlinearities in the scaling factors, making it a very versatile model to choose in hyperspectral image unmixing when nonlinearities and endmember variability are significant.

We refer the interested reader to~\cite{drumetztip} for detailed descriptions of algorithms which are able to estimate the parameters of both versions of the ELMM (abundances and scaling factors).
\section{Experimental validation}
\label{exp}
In this section, we provide a qualitative and quantitative analysis of the quality of the approximations of the full Hapke model necessary to reach the model of~\eqref{scaling}. The goal is not to evaluate the relevance and performance of the ELMM in unmixing applications, which has already been extensively carried out in~\cite{drumetztip, Veganzones2014_ELMM, drumetz2017relationships}. For the three endmembers of Fig.~\ref{pca2}, for which we obtained estimates of the albedo spectra and the photometric parameters~\cite{marrerovalidation, cord2005experimental}, we compare the reflectance endmembers generated in several configurations.

\begin{figure}
\includegraphics[scale=0.203]{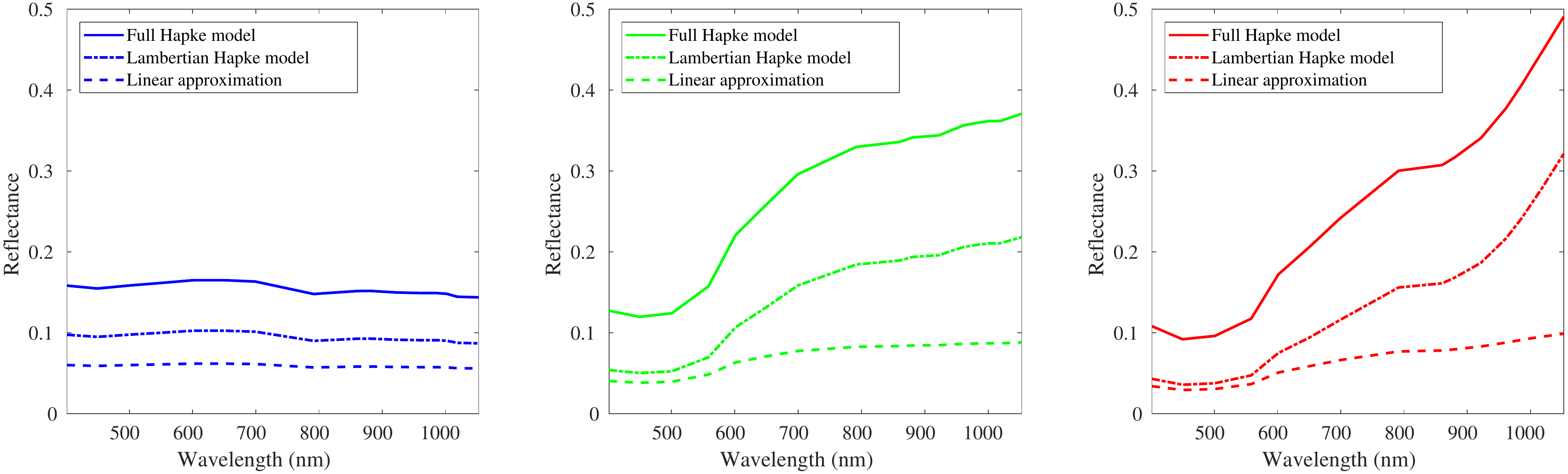}\\
\includegraphics[scale=0.2025]{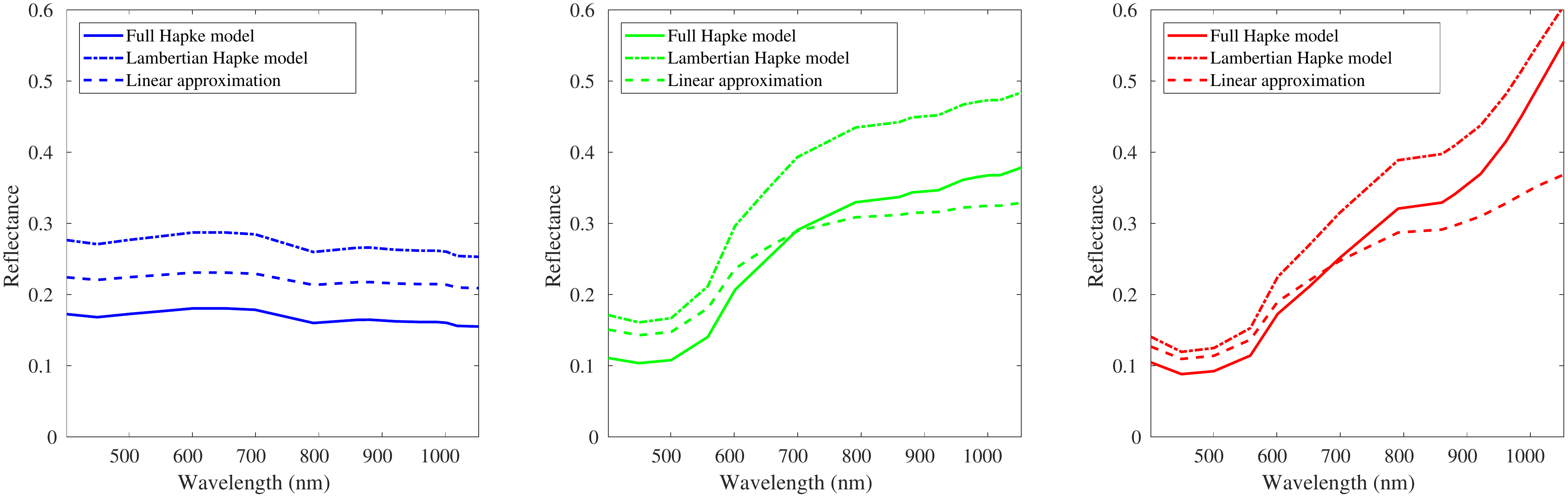}
\caption{Reflectances computed from the albedos, the photometric and geometric parameters using the full Hapke Model~(\ref{full_hapke}), the Lambertian approximation~\eqref{simple_hapke} and the linear approximation~\eqref{Taylor}. The columns correspond to the materials; from left to right: basalt (blue), palagonite (green) and tephra (red). The rows correspond to two different angular configurations (top: $\theta_{0} = 0^{\circ}$ and $\theta = 0^{\circ}$, $\phi = 0^{\circ}$, bottom: $\theta_{0} = 90^{\circ}$ and $\theta = 45^{\circ}$, $\phi = 0^{\circ}$).}
\label{refl_nadir_raking}
\end{figure}

\begin{figure}
\vspace{-0.5cm}
\includegraphics[scale=0.21]{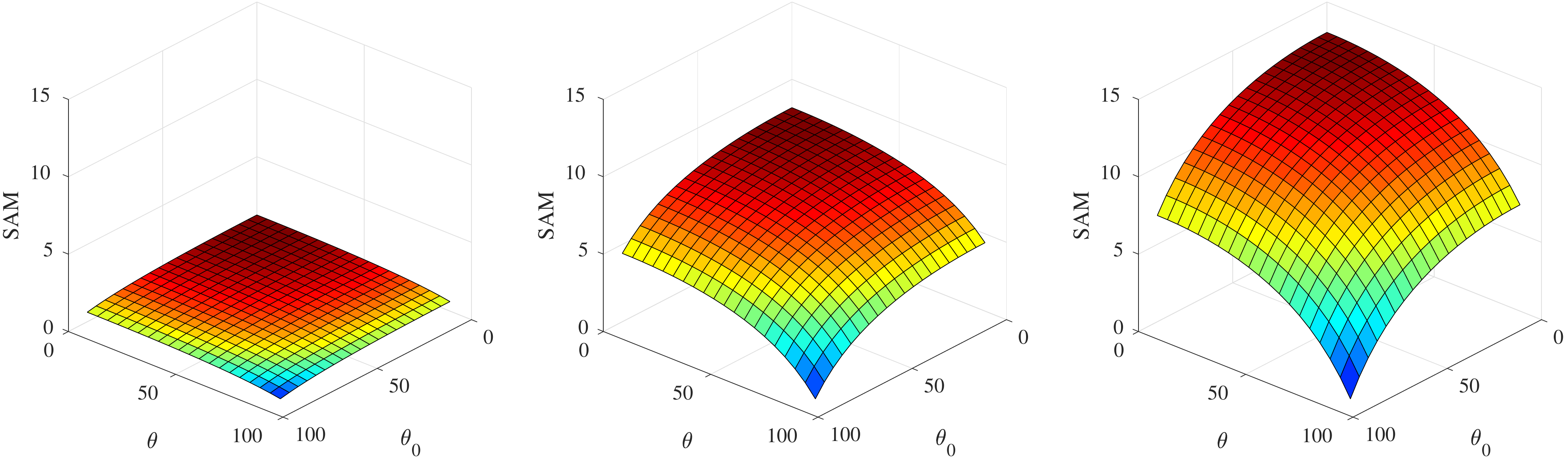}\\
\includegraphics[scale=0.2]{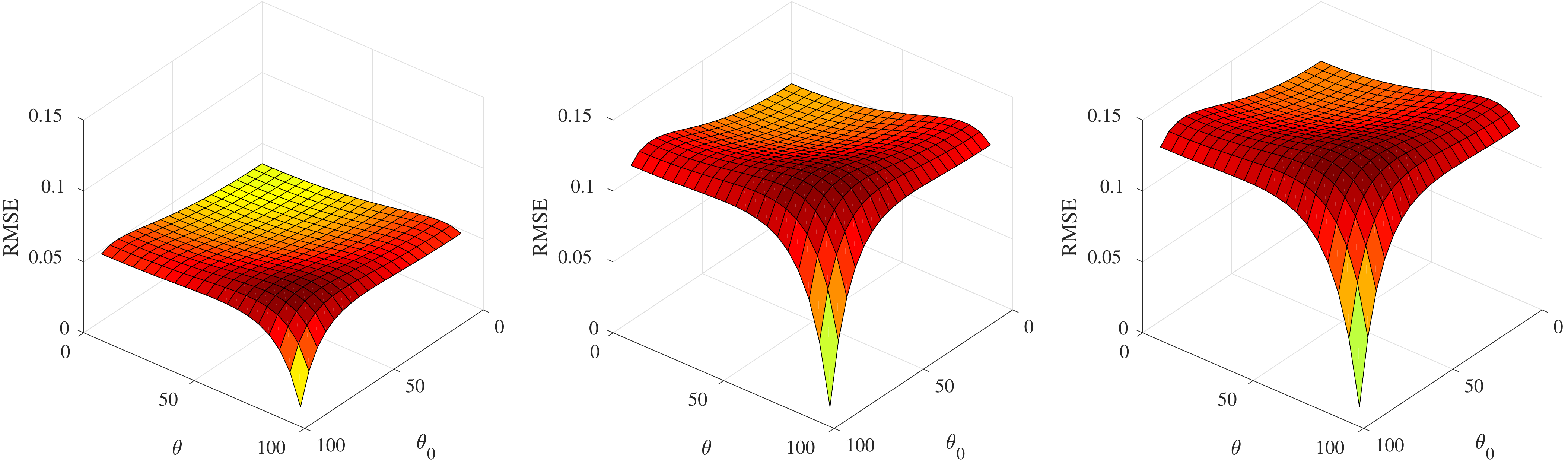}
\caption{Spectral angle (top row) and Root Mean Squared Errors (bottom) between the reflectance spectra generated by the Lambertian model~\eqref{simple_hapke} and the linear approximation~\eqref{Taylor} plotted against the incidence angle $\theta_0 $ and the emergence angle $\theta$ (in degrees), for each material (from left to right: basalt, palagonite and tephra).}
\label{sam_rmse}
\end{figure}

In Fig.~\ref{refl_nadir_raking}, we show how reflectance is calculated from the full Hapke Model~\eqref{full_hapke}, the Lambertian approximation~\eqref{simple_hapke} and the linear approximation~\eqref{Taylor}, for basalt, palagonite and tephra, whose photometric parameters are known~\cite{cord2005experimental}. We consider two geometries corresponding to those of Fig.~\ref{albedos} (a) ($\theta_{0} = \theta = 0^{\circ}$) and (c) ($\theta_{0} = 90$ and $\theta = 45^{\circ}$). In the first configuration, we see that the Lambertian approximation and the linear approximation are quite coarse, but acceptable for small reflectance values, while important discrepancies appear for larger reflectance values. The second geometrical configuration leads to a better approximation and a better agreement between the Lambertian model and the linear approximation since the relationship between albedo and reflectance becomes more and more linear as the angles get closer to $90^{\circ}$.

For a more thorough analysis of the accuracy of the linear approximation depending on the geometry, we show in Fig.~\ref{sam_rmse} plots of the spectral angle and the root mean squared errors (RMSE) between the reflectance spectral generated by the Lambertian model and the linear approximation for the three materials of interest for various angular configurations. There is a perfect agreement between both models when $\theta_{0} = 90$ and $\theta = 90^{\circ}$, and the quality of the approximation decreases as the angles become smaller. Interestingly, the curvature of the spectral angle surface seems to be directly related to the average albedo: the approximation is always satisfactory for basalt, while it is coarser for palagonite and even coarser for tephra. RMSE is overall less affected by the geometry, but the RMSE level is again directly related to the average albedo.
\section{Conclusion}
\label{concl}
In the hyperspectral image processing community, it has long been known empirically that scaling factors can efficiently and conveniently model brightness variations due to changing illumination conditions. In this letter, we have theoretically connected the Extended Linear Mixing Model, a tractable model taking explicitly this phenomenon into account for spectral unmixing and the the semi-empirical Hapke model by making simplifying assumptions. We prove and experimentally verify that these assumptions are the most reasonable when the albedo is not too large, or for favorable geometric configurations. Combined with the capability of the ELMM to locally approximate nonlinear mixtures~\cite{drumetz2017relationships}, this result further motivates the use of the ELMM to unmix images in which nonlinearities and/or variability effects are non negligible.
\bibliographystyle{ieeetr}

\bibliography{references}
\vspace{-1cm}
\end{document}